\newif\ifconf
\conftrue
\ifconf
	\documentclass[conference, lettersize]{IEEEtran}	
\else
	\documentclass[journal]{IEEEtran}	
\fi

\usepackage{amsmath,mathtools,amssymb}
\usepackage[svgnames]{xcolor}
\usepackage{tikz,pgfplots}
\usetikzlibrary{calc,arrows,shapes,backgrounds,calc,positioning,patterns,decorations.pathmorphing,decorations.markings,mindmap,trees, arrows.meta, positioning}
\usepackage{amsthm} 
\usepackage{blox}
\usepackage{color}
\usepackage{siunitx}	
\usepackage{booktabs}
\usepackage{balance} 
\usepackage{algorithm}
\usepackage{algpseudocode}
\usepackage{cite}
\usepackage{dsfont}
\usepackage[font=small]{caption}
\usepackage[dvipsnames]{xcolor}
\usepackage{subcaption}
\usepackage{comment}
\usepackage{enumitem}
\DeclareFontFamily{U}{mathx}{\hyphenchar\font45}
\DeclareFontShape{U}{mathx}{m}{n}{<-> mathx10}{}
\DeclareSymbolFont{mathx}{U}{mathx}{m}{n}
\DeclareMathAccent{\widebar}{0}{mathx}{"73}

\newtheorem{remark}{Remark}
\newtheorem{assumption}{Assumption}

\makeatletter
\xdef\@endgadget#1{{\unskip\nobreak\hfil\penalty50\hskip1em\hbox{}\nobreak\hfil#1\parfillskip=0pt\finalhyphendemerits=0\par}}
\newcommand\@Endofsymbol{$\triangledown$}
\newcommand\Endofremark{\@endgadget{\@Endofsymbol}}
\makeatother

\newcommand{\R}{\mathbb{R}}

\newcommand{\N}{\mathbb{N}}

\newcommand{\interior}[1]{%
	{\kern0pt#1}^{\mathrm{o}}%
}

\definecolor{BlockDiagramGreen}{HTML}{CCF6C8} 
\definecolor{BlockDiagramBlue}{HTML}{BEEBE9} 
\definecolor{BlockDiagramYellow}{HTML}{FFF2CC}
\definecolor{BlockDiagramRed}{HTML}{FF8787}

\DeclareMathOperator*{\argmin}{\arg\!\min}

\algdef{SE}[EVERY]{Every}{EndEvery}[1]{\textbf{Every} #1 \textbf{do}}{\textbf{end every}}

\title{Optimal Dispatch of Connected and Autonomous Electric Vehicles to Enhance Short-Term Grid Flexibility in Smart Cities}

\author{
Nikolas Sacchi, Giacomo Basile, Silvia Siri, Manuela Minetti, Andrea Bonfiglio, Antonella Ferrara 
	\thanks{
    \newline
		\indent Nikolas Sacchi and Antonella Ferrara are with the Department of Electrical,
		Computer and Biomedical Engineering, University of Pavia, Pavia, Italy (e-mail: \textsl{nikolas.sacchi01@universitadipavia.it}, \textsl{antonella.ferrara@unipv.it}).
       \newline 
       \indent Giacomo Basile and Silvia Siri are with the Department of Computer Science, Bioengineering, Robotics and Systems Engineering, University of Genoa, Genoa, Italy (e-mail: \textsl{giacomo.basile@edu.unige.it }, \textsl{silvia.siri@unige.it}). 
       \newline 
       \indent Manuela Minetti and Andrea Bonfiglio are with the Department of Electrical, Electronic, Telecommunications Engineering and Naval Architecture, University of Genoa, Genoa, Italy (e-mail: \textsl{manuela.minetti@unige.it}, \textsl{a.bonfiglio@unige.it}). 
    \newline Work developed in the frame of “E-COSMOS – Electric Charging Optimization for Sustainable Mobility Systems” project financed by the Italian Ministry of University and Research under the Call for Joint Research and Innovation Projects 2024-2026 Between Italy and the Republic of Serbia. CUP: D33C25000440001, F13C25000270001.
}}

\IEEEoverridecommandlockouts     
\begin{document}
	\maketitle
	\thispagestyle{empty}
	
	
	\begin{abstract}
		This paper proposes a coordinated energy–mobility dispatch framework for grid support service provision in smart cities under time constraints. In particular, a scenario  in which a distributed system operator requests a specified amount of energy within a given deadline is considered. A fleet of connected autonomous electric vehicles equipped with virtual battery partitioning is dynamically dispatched toward vehicle-to-grid stations. The routing problem is formulated as a periodically updated resource-constrained shortest path, accounting for time and energy constraints with congestion-dependent travel times derived from a dynamic traffic model. At the vehicle level, a model predictive control strategy regulates speed to satisfy mobility energy requirements while ensuring deadline compliance. The framework is validated through simulations on the urban network of Rapallo (Italy), demonstrating robustness against congestion-induced delays.
	\end{abstract}
	\textbf{\textit{Index terms -- } Connected autonomous electric vehicles, optimal traffic-aware dispatch, power grid flexibility services, vehicle-to-grid solutions.}
		

    \section{Introduction}

    The rapid advancement of communication, sensing, and control technologies is progressively reshaping urban environments into interconnected data-driven ecosystems, referred to as \textit{smart cities}~\cite{rua2026smart}, in which urban infrastructures are conceived as interoperable layers where communication networks, energy and mobility are deeply integrated~\cite{zanella2014internet}. Thus, coordinated management of heterogeneous resources is promoted to enhance sustainability and operational efficiency at urban scale~\cite{bibri2017smart, jia2022control}.
 
     In this context, the energy sector is undergoing a profound transition driven by the increasing penetration of Renewable Energy Sources (RESs). Modern smart grids, enabled by advanced sensing, communication, and distributed control technologies, allow Distributed System Operators (DSOs) to coordinate Distributed Energy Resources (DERs) in real time~\cite{akorede2010distributed}, thereby facilitating large-scale RES integration~\cite{twidell2021renewable}. However, the inherent intermittency of renewable generation introduces operational challenges, including supply–demand imbalances, voltage deviations, and frequency fluctuations. Consequently, the provision of Grid Support Services (GSSs), such as frequency regulation and voltage control~\cite{de2025exploring}, has become central to contemporary power systems, increasingly relying on distributed and fast-acting urban assets to ensure flexibility and grid reliability, under high RES penetration.

     In parallel with the growing need for distributed flexibility, the transportation sector is rapidly electrifying, and the increasingly present Electric Vehicles (EVs) and Connected Autonomous EVs (CAEVs) are introducing a large pool of battery-based distributed storage resources into urban systems~\cite{torkey2024transportation}. Through the Vehicle-to-Grid (V2G) paradigm, these vehicles enable bidirectional power exchange with the grid, allowing parked EVs to provide grid support services such as demand response, peak shaving, frequency regulation, and voltage control~\cite{abdulaal2016solving,vstogl2024electric}. 
     

     In order to coordinate EV participation in GSSs, several control and optimization strategies have been proposed. For example, offline optimal dispatch approaches include coordinated active and reactive power control for demand-side management~\cite{wang2022coordinated}, day-ahead scheduling strategies for peak shaving through controlled EV discharging~\cite{dong2023multi}, and hierarchical architectures jointly optimizing power flow and V2G scheduling to address frequency regulation~\cite{zhang2020joint}. To enhance adaptability to real-time grid conditions, some works (see, e.g., \cite{hu2021distributed}) exploit distributed Model Predictive Control (MPC) schemes for voltage regulation using V2G-enabled chargers.

    Despite these advancements, large-scale V2G deployment remains limited by technical, economic, and social constraints, including battery degradation, uncertainty in vehicle availability, and stochastic user behavior, which undermine the predictability of aggregated flexibility~\cite{mavlutova2023urban}. From a technical perspective,  uncertainties related to battery degradation can be mitigated through the Virtual Battery Partitioning (VBP) proposed in~\cite{bonfiglio2025vehicle}, which virtually separates the onboard battery into mobility and GSS layers, preserving driving requirements while enabling reliable service provision. From an organizational standpoint, the diffusion of car sharing systems, which act as fleet aggregators with centralized management, enables coordinated scheduling and improved controllability of EVs (or CAEVs) availability, thereby enhancing the reliability of GSSs~\cite{mavlutova2023urban}.
    
    Nevertheless, existing methodologies generally neglect the interaction between energy service provision and traffic dynamics, overlooking mobility-induced uncertainties such as congestion-related delays that may compromise the timely delivery of grid support services in real urban environments.

    To address this gap, this paper proposes a coordinated energy–mobility dispatch framework for grid support services in smart cities. In particular, it considers a scenario in which the DSO requests a prescribed amount of energy to be delivered within a given time horizon by means of a fleet of CAEVs equipped with VBP which is dynamically dispatched toward V2G stations.
    The proposed architecture adopts a two-layer decision scheme. At the upper layer, vehicle routing is determined by periodically solving a Resource-Constrained Shortest Path (RCSP) problem over the urban network, where time and energy are treated as constrained resources to guarantee feasibility with respect to both the DSO deadline and the available grid-support energy. Routing costs are updated measuring the state of a dynamic traffic model that emulates a real traffic network, enabling congestion-aware and adaptive dispatch decisions. At the lower layer, an MPC strategy regulates each vehicle’s speed along the assigned path, enforcing mobility energy constraints while ensuring arrival within the required time window. The framework has been validated through numerical simulations on the real urban network of Rapallo (Italy), where traffic dynamics is described by the macroscopic model in~\cite{dinopoulou2006applications}.

    The remainder of the paper is organized as follows. Section~\ref{sec:preliminaries} introduces the preliminary concepts and formalizes the problem formulation. Section~\ref{sec:proposal} details the proposed framework. Simulation results are discussed in Section~\ref{sec:numerical_validation}, and concluding remarks are provided in Section~\ref{sec:conclusion}.

    \section{Preliminaries and Problem Formulation} \label{sec:preliminaries}
    This section introduces the urban traffic network modeling, the virtual battery partitioning paradigm, and formally defines the problem that is solved in this paper.

    \subsection{Urban Network Modeling}\label{sec:urban_network_modeling}

    We consider an urban traffic network composed of \( J \in \N_{>0} \) road junctions connected by \( M \in \N_{>0} \) directed road links. In addition, \( N_{\mathrm{V2G}} \in \N_{>0} \) bidirectional V2G stations are deployed within the network in some specific junctions (e.g., in dedicated areas such as parking facilities). The network is represented as a directed graph \( \mathcal{G} = (\mathcal{V}, \mathcal{E}) \), where \( \mathcal{V} \) denotes the set of nodes corresponding to road junctions, characterized by cardinality \( |\mathcal{V}| = J \), and \( \mathcal{E} \subseteq \mathcal{V} \times \mathcal{V} \) represents the set of directed road links, with \(|\mathcal{E}| = M\). The nodes equipped with V2G stations are identified by the set \(\mathcal{V}_{\mathrm{V2G}} \subset \mathcal{V}\), with \( |\mathcal{V}_{\mathrm{V2G}}| = N_{\mathrm{V2G}} \). The urban area is assumed to be partitioned into \( D \in \N_{>0} \) energy districts, collected in the set \( \mathcal{D} \), with \( |\mathcal{D}| = D \). For each district \( d \in \mathcal{D} \), we define the subset \( \mathcal{V}_{\mathrm{V2G}, d} \subseteq \mathcal{V}_{\mathrm{V2G}} \) of V2G nodes belonging to district \( d \), such that \( \bigcup_{d \in \mathcal{D}} \mathcal{V}_{\mathrm{V2G}, d} = \mathcal{V}_{\mathrm{V2G}}\).

	\subsection{Urban Traffic Dynamics} \label{sec:urban_traffic_dynamics}
   In the following, we present the mathematical formulation of the urban traffic dynamics adopted in the simulations, which is the one proposed in \cite{dinopoulou2006applications}.
    

    For each junction \(j \in \mathcal{V}\), we define the sets of incoming and outgoing links as \(\mathcal{I}_{j} \subset \mathcal{E}\) and \(\mathcal{O}_{j} \subset \mathcal{E}\), respectively. Moreover, we define a map  \(\xi_{j} : \mathcal{I}_{j} \times \mathcal{O}_{j} \rightarrow [0, 1]\), for the flow leaving the links in \(\mathcal{I}_{j}\) toward the links \(\mathcal{O}_{j}\).

    \begin{remark}[Choice of \(\xi_{j}\)]
        \label{rmk:turning_rates}
         In this work, the map $\xi_{j}$ is defined such that drivers tend to choose paths leading toward a subset \(\mathcal{T}_{\mathrm{sink}} \subset \mathcal{T}\), where \(\mathcal{T} \subset \mathcal{V}\) represent the set of terminal junctions, i.e., dead ends or boundary of the graph toward extra-urban roads. Given an incoming link  $e_{in} \in \mathcal{I}_{j}$, the fraction of flow directed toward an outgoing  link $e_{out} = (j, w) \in \mathcal{O}_{j}$ is weighted by the shortest path distance from $w$ to the nearest sink, so that outgoing links leading to nodes closer to $\mathcal{T}_{\mathrm{sink}}$ receive a proportionally larger share of the flow. Finally, in order to avoid u-turns, pairs $(e_{in}, e_{out})$ such that $e_{out}$ leads back to the upstream node of $e_{in}$, are explicitly forbidden by setting $\xi_{j}(e_{in}, e_{out}) = 0$.
    \end{remark}
    
    In this paper, we consider the junction set \(\mathcal{V}\) to be subdivided into two subsets, i.e., \(\mathcal{V} = \mathcal{V}_{\mathrm{ROW}} \cup \mathcal{V}_{\mathrm{TL}}\), where the former consists of normal junctions where the vehicles, coming from the incoming links, have always the right of way (ROW), while the latter accounts for junctions governed by traffic lights (TLs). Each junction \(j \in \mathcal{V}_{\mathrm{TL}}\) is characterized by a cycle time [h] and total lost time [h] denoted as \(H_{j} \in \R_{> 0}\) and \(C_{j} \in \R_{> 0}\), respectively. Moreover, each signalized junction operates with a finite number of signal stages collected in the set \( \mathcal{S}_{j} \). Each stage \( s \in \mathcal{S}_{j} \) corresponds to a subset of incoming links \( \mathcal{I}_{j} \) that are granted the ROW during that stage, and it is characterized by an effective green time [h], denoted by \( g_{s,j} \in \R_{>0} \). Without loss of generality, the following assumption on the cycle, lost, and green times is adopted. Note that, for each junction \( j \in \mathcal{V}_{\mathrm{TL}} \), the cycle time and the total lost time are assumed to be fixed and identical across junctions, i.e., \( C_{j} = C \) and \( H_{j} = H \), where \( C, H \in \R_{>0} \) are given constants. Moreover, for all stages \( s \in \mathcal{S}_{j} \) and all junctions \( j \in \mathcal{V}_{\mathrm{TL}} \), the effective green times are assumed to be known and identical, namely \( g_{s,j} = \bar{g} \), with \( \bar{g} \in \R_{>0} \).

    Consider link \((i,h)\) leaving junction \(i \in \mathcal{V}\) toward  junction \(h \in \mathcal{V}\). We denote all quantities associated with that link with subscript \(i\) and superscript \(h\). Then, the number of vehicles in it is denoted as \(x_{i}^{h} \in \N_{\geq 0}\) and is characterized by dynamics 
    \begin{equation}\label{eq:link_dyn}
        x_{i}^{h}(k+1) = x_{i}^{h}(k)+T\Big(q_{i}^{h}(k)-s_{i}^{h}(k)+d_{i}^{h}(k)-l_{i}^{h}(k)\Big),    
        \end{equation}
    where \(q_{i}^{h} \in \R_{\geq 0}\) is the flow [veh/h] entering the link from junction \(i\), \(l_{i}^{h} \in \R_{\geq 0}\) is the flow [veh/h] leaving toward junction \(h\), \(s_{i}^{h} \in \R_{\geq 0}\) represents the flow [veh/h] that leaves the link without being injected into junction \(h\), while \(d_{i}^{h} \in \R_{\geq 0}\) represents the flow [veh/h] that enters the link from a source that is not junction \(i\), and it represents a boundary condition. As for \(k \in \N_{\geq 0}\) and \(T \in \R_{>0}\), they denote the time instant and the sampling time [h].  

    The flow \(q_{i}^{h}\) depends on the flows leaving the links \(\mathcal{I}_{i}\) and on the associated turning rates toward link \((i,h)\). In particular,
    \begin{equation}\label{eq:inflow_q}
        q_{i}^{h}(k)=\sum_{(u,i)\in\mathcal{I}_i} \xi_{i}\big((u,i), (i,h)\big)\,l_u^i(k).
    \end{equation}
    As for the exit flow \(s_{i}^{h}\), it is computed as a partition of the flow in \eqref{eq:inflow_q}, i.e., 
    \begin{equation}
        \label{eq:exit_flow}
        s_{i}^{h}(k)=\epsilon_{i}^{h}\,q_{i}^{h}(k),
    \end{equation}
    where $\epsilon_{i}^{h} \in \R_{>0}$ is a constant representing the exit rate. 

    The computation of the flow \( l_{i}^{h} \) depends on the type of junction \( h \). In particular, if \( h \in \mathcal{V}_{\mathrm{ROW}} \), i.e., the junction is not signalized, then the incoming link \( (i,h) \) is permanently granted the ROW, and the flow is given by
    \begin{equation}
        \label{eq:out_flow_row}
        l_{i}^{h}(k) = \min\left\{ \frac{x_{i}^{h}(k)}{T}, \, \Phi_{i}^{h} \right\},
    \end{equation}
    where \( \Phi_{i}^{h} \in \R_{\geq 0} \) denotes the saturation flow rate [veh/h], which depends on the geometric and operational characteristics of the link.

    On the other hand, if \(h \in \mathcal{V}_{\mathrm{TL}}\), then the flow \(l_{i}^{h}\) is non zero only when link \((i,h)\) has ROW. Specifically, Let \(\mathcal{S}_{h,i} \subseteq \mathcal{S}_{h}\) be the subset of stages in which link \((i,h)\) has ROW. Then, choosing \(T > C\) \cite{dinopoulou2006applications}, the flow \(l_{i}^{h}\) is given by \(l_{i}^{h}(k) = \min \left\{ \frac{x_{i}^{h}(k)}{T}, \tfrac{\Phi_{i}^{h}\,|\mathcal{S}_h^{i}| \bar{g}}{C}\right\}\).
    

	\subsection{CAEV Virtual Battery Partitioning}
    \label{sec:virtual_partitioning}
    
    In this paper, we assume that each CAEV is equipped with a Battery Management System (BMS) capable of dividing the vehicle battery into two virtual partitions ~\cite{bonfiglio2025vehicle}: one dedicated to vehicle mobility and one allocated to GSSs. 
    
    Considering a generic CAEV \( c \), the two virtual batteries are characterized by maximum virtual capacities denoted by \( \bar{E}_{c,\mathrm{mob}} \in \R_{\geq 0} \) and \( \bar{E}_{c,\mathrm{GSS}} \in \R_{\geq 0} \), respectively. These quantities are defined as fractions of the maximum capacity of the physical battery \( \bar{E}_{c} \in \R_{>0} \), namely \(\bar{E}_{c,\mathrm{GSS}} = \alpha_{c, \mathrm{GSS}} \bar{E}_{c}\) and \(\bar{E}_{c,\mathrm{mob}} = (1 - \alpha_{c,\mathrm{GSS}}) \bar{E}_{c}\), 
    where \( \alpha_{c,\mathrm{GSS}} \in [0,1] \) denotes the constant virtual partition factor. The actual energy levels, assumed to be measurable and denoted as \( E_{c,\mathrm{GSS}} \in [0,\bar{E}_{c,\mathrm{GSS}}] \) and \( E_{c,\mathrm{mob}} \in [0,\bar{E}_{c,\mathrm{mob}}] \), are characterized by discharge dynamics
    \begin{equation}
        \label{eq:energy_dynamics}
        E_{c,\mathrm{mob}}(k+1) = E_{c,\mathrm{mob}}(k) - \big(\eta_1 u_c(k) + \eta_2 u_c(k)^2\big) \, T
    \end{equation}
    where \(u_c(t) \in \R_{\ge 0}\) is the vehicle speed, while \(\eta_1, \eta_2 \in \R_{> 0}\) are coefficients capturing rolling resistance and aerodynamic drag \cite{rajamani2006vehicle}.
    
    It is important to emphasize that this partitioning does not imply any physical separation within the battery pack. Rather, as detailed in \cite{bonfiglio2025vehicle}, it is implemented at the energy management level through software-defined constraints.
   

	\subsection{Problem Formulation} \label{sec:problem_formulation}
    According to the urban network modeling in Section \ref{sec:urban_network_modeling}, each energy district \( d \in \mathcal{D} \) is managed by the DSO and embeds DERs, such as photovoltaic units, small wind turbines, and battery storage systems. Due to the variability and intermittency of DER generation, temporary imbalances between local production and consumption may arise. To cope with the induced imbalances, each energy district is assumed to be equipped with short-term forecasting tools that provide, for every district \( d \in \mathcal{D} \), an estimate of the energy required over a future time horizon \([t_{\mathrm{pr}},\, t_{\mathrm{pr}} + T_{\mathrm{pr}}]\) to ensure GSS. This quantity is denoted by \( \hat{E}^{\star}_{d,\mathrm{GSS}} \in \R_{\geq 0}\). 

    Consider a set of CAEVs \( \mathcal{C} \), each equipped with a BMS enabling the VBP described in Section~\ref{sec:virtual_partitioning}. The objective of this paper is to propose a control architecture that, given the forecasted energy requirement \( \hat{E}^{\star}_{d,\mathrm{GSS}} \), coordinates and dispatches CAEVs toward the V2G nodes \( \mathcal{V}_{\mathrm{V2G}, d} \) of the corresponding district. Let \( t_{\mathrm{disp}} < t_{\mathrm{pr}} \) denote the instant at which the dispatching procedure is initiated. The dispatch must be completed before \( t_{\mathrm{pr}} \), i.e., before the beginning of the horizon in which the energy support is required.
    Furthermore, the allocation of vehicles must respect both travel-time constraints and the available mobility energy \(E_{c, \mathrm{mob}}\) of each CAEV \(c\) involved in the dispatch.

			
			
			
			
			

	\section{The Dispatch Control Architecture} \label{sec:proposal}
     \begin{figure}[t]
	   \centering
	   \begin{tikzpicture}[scale=1]
	   		\node [draw, align=center, font=\footnotesize ]  (urban_network) at (0,0){ Urban Network };
	   
        \node[above right = 0.1 pt and 0.1 pt of urban_network.east, align=center, font = \scriptsize](){Load data};
        \node[below right = 0.1 pt and 0.1 pt of urban_network.south, align=center, font = \scriptsize](){Real-time \\ Traffic data};
        \node[above left=0.1pt and 0.1pt of urban_network.west](){\scriptsize{Fleet Speed}};
        
         \node[draw, below right=1 cm and 0.9 cm of urban_network, align=center, fill=White, font=\footnotesize] (dso) {
	   		\begin{minipage}[t][1.2cm][t]{1.5cm} \centering DSO \(d \in \mathcal {D}\) 	\end{minipage}};
        
         \node[draw=black, fill=white, minimum width=0.5cm, above=0.15cm of dso.south, align=center, font=\footnotesize] {Energy \\ Forecaster};
          \node[above left = 0.1 pt and 0.1 pt of dso.west, align=center](){\scriptsize{\(\hat{E}^{\star}_{d, \mathrm{GSS}}\)}};
          \node[left = 0.5 cm of dso.north](dso_left){};

         \node[draw,  left= 1 cm of dso, align=center, fill=White] (aggregator) {
	   			\begin{minipage}[t][1.5cm][t]{1.55cm} 
 	   				\centering
	   				\footnotesize{CAEV Aggregator} 	
	   			\end{minipage}
	   		};
        \node[draw=black, fill=BlockDiagramYellow, minimum width=0.5cm, above=0.15cm of aggregator.south, font=\footnotesize, align=center] {Traffic-Aware \\ Path Planner};
        \node[above left = 0.1 pt and 0.1 pt of aggregator.west, align=center, font=\scriptsize](){Optimal \\ paths};
        \node[left=0.25cm of aggregator.north](fleet_mid_3){};

        \node[draw, dashed, inner sep = 1pt, left= 1 cm of aggregator, minimum width = 3 cm, align=center, minimum height=1.4cm, fill=White] (fleet) {};
        \node[below = 0.1pt of fleet.south, font=\scriptsize](){CAEV Fleet};
        \node[right=0.1cm of fleet.north](fleet_mid_1){};
        \node[above=0.35cm of fleet_mid_1.center](fleet_mid_2){};
        \node[above right=0.1pt and 0.1pt of fleet_mid_2.center](){\scriptsize{Fleet Status}};

        \node[draw=black, fill=white, inner sep = 2pt, minimum width=0.5cm, below left =0.15cm and 0.2cm of fleet.north] (cav1) {
            \begin{minipage}[t][1.cm][t]{1cm} 
 	   				\centering
	   				\scriptsize{CAEV 1} 	
	   		\end{minipage}
        
            };
        \node[draw=black, fill=BlockDiagramRed, minimum width=0.5cm, above=0.25cm of cav1.south, font=\scriptsize] {MPC};
        
        \node[inner sep = 0 pt, outer sep = 0pt, right =0.05cm of cav1.east] (dots) {\footnotesize{\dots}};
        \node[draw=black, fill=white, inner sep = 2pt, minimum width=0.5cm, right =0.05cm  of dots.east] (cav2) {\begin{minipage}[t][1.cm][t]{1cm} 
                    \centering
                    \scriptsize{CAEV C} 	
            \end{minipage}
            };
        
        \node[draw=black, fill=BlockDiagramRed, minimum width=0.5cm, above=0.25cm of cav2.south, font=\scriptsize] {MPC};
          \node [draw, inner sep=0pt, above=0.25 cm of urban_network](solar){ \includegraphics[width=25px]{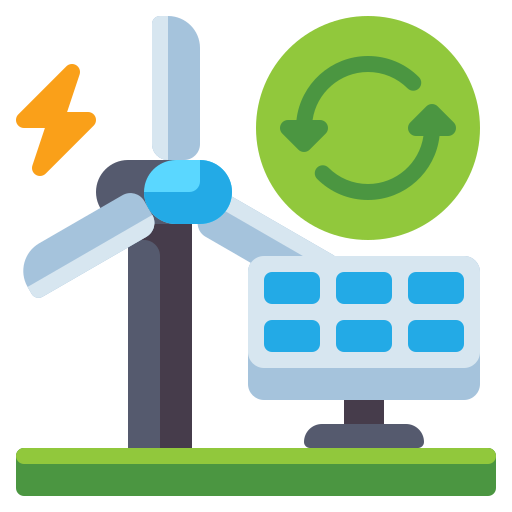}};
           \node[above right = 0.1 pt and 0.1 pt of solar.east, align=left, font=\scriptsize](){Production data};
           
	 		\draw[-stealth] (urban_network.east) -| (dso_left.center); 
         \draw[-stealth] (dso.west) -- (aggregator.east); 
           \draw[-stealth] (solar.east) -| (dso.north); 
           \draw[-stealth] (aggregator.west) -- (fleet.east); 
           \draw[-stealth] (fleet.north) |- (urban_network.west); 
            \draw[-stealth] (urban_network.south) -- (aggregator.north); 
        \draw[-] (fleet_mid_1.center) -- (fleet_mid_2.center); 
        \draw[-stealth] (fleet_mid_2.center) -| (fleet_mid_3.center);

		\end{tikzpicture}
        \caption{Block diagram of the  architecture, in which the proposed path planner and the MPC are highlighted in yellow and red.}
		
		\label{fig:block_diagram}
	\end{figure}
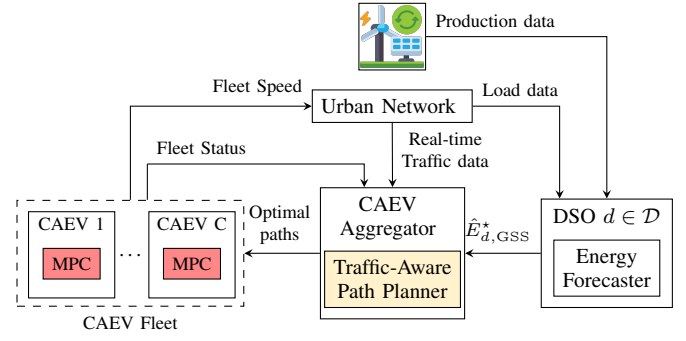  

    The aim of this section is to present the proposed control architecture, which is  schematized in Fig. \ref{fig:block_diagram} and composed by two main components: a \textit{traffic-aware path planner} and a \textit{CAEV-level MPC scheme}. 

    \subsection{Traffic-Aware Path Planner}
    Consider a DSO that, at time \( t_{\mathrm{disp}} \), issues an energy request \( \hat{E}^{\star}_{d, \mathrm{GSS}} \) for district \( d \in \mathcal{D} \), referring to the horizon \( [t_{\mathrm{pr}},\, t_{\mathrm{pr}} + T_{\mathrm{pr}}] \). Let \( \bar{\mathcal{C}} \subseteq \mathcal{C} \) denote the subset of CAEVs satisfying the following assumption.
    \begin{assumption}
        \label{ass:cav_availability}
        Each CAEV \( c \in \bar{\mathcal{C}} \) is not scheduled for mobility tasks over the interval \( [t_{\mathrm{disp}},\, t_{\mathrm{pr}} + T_{\mathrm{pr}}] \).
    \end{assumption}

    Consider now a subset \( \bar{\mathcal{C}}_{d} \subseteq \bar{\mathcal{C}}\) of CAEVs identified for GSS of district \(d\). 
    The objective of the planner is to determine, at a given time instant \( t_{k} \in [t_{\mathrm{disp}}, t_{\mathrm{pr}}] \) and for each CAEV \( c \in \bar{\mathcal{C}}_{d} \), the optimal target node, denoted as \(n^{\star}_{c, \mathrm{V2G}} \in \mathcal{V}_{\mathrm{V2G}}\), and the associated optimal path. This routing problem can be computed by solving an RCSP problem. 

    Let \(n_{c} \in \mathcal{V}\) be the location of CAEV \(c\) at time \(t_{k}\) and let \(\bar{\mathcal{P}}_c(n_{\mathrm{V2G}})\) denote the set of all paths that connect node \(n_{c}\) and a generic V2G node \(n_{\mathrm{V2G}}\), where each path is defined as a sequence of edges. Then, the optimal path  \(\mathcal{P}^{\star}_c(n_{\mathrm{V2G}}) \in \bar{\mathcal{P}}_c(n_{\mathrm{V2G}})\) considering the traffic condition a time \(t_{k}\) is obtained by solving the following optimal problem
    \begin{subequations}
        \label{eq:rcsp}
        \begin{align}
            \mathcal{P}^\star_c(n_{\mathrm{V2G}}) = \argmin_{p \in \bar{\mathcal{P}}_c(n_{\mathrm{V2G}})} & J(p) \coloneqq \sum_{(i,h) \in p} \tau_{i}^{h}(t_{k}) \label{eq:rcsp:cost}\\
            \text{s.t.} \quad 
            &\sum_{(i,h) \in p} \tau_{i}^{h}(t_{k})\leq \Delta_{\mathrm{disp}} - \Delta_{k}, \label{eq:rcsp:time_constraint}\\
            &\sum_{(i,h) \in p} \varepsilon_{i}^{h} \leq E_{c,\mathrm{mob}}(t_k)\label{eq:rcsp:energy_constraint},
        \end{align}
    \end{subequations}
    where \(\Delta_{\mathrm{disp}}=t_{\mathrm{pr}}-t_{\mathrm{disp}}\) is the maximum dispatch time, \(\Delta_{k} = t_{k} - t_{\mathrm{disp}}\), while  \(\tau_{i}^{h} \in \R_{>0}\) and \(\varepsilon_{i}^{h} \in \R_{>0}\) are the travel time [h] and distance-based energy consumption [kWh] for link \((i,h)\). Specifically, \eqref{eq:rcsp:cost} allows to minimize the travel time between \(n_{c}\) and \(n_{\mathrm{V2G}}\), \eqref{eq:rcsp:time_constraint} ensures that the path is completed before the deadline \(t_{\mathrm{pr}}\), while \eqref{eq:rcsp:energy_constraint} ensures that the CAEV has sufficiently mobility energy to complete the path.
    

    If, on the one hand, the term \(\varepsilon_{i}^{h}\) is constant and dependent on the shape of the road link, on the other \(\tau_{i}^{h}\) is strictly related to the traffic conditions,  depending on the number of vehicles present in the link, its length \(\lambda_{i}^{h} \in \R_{>0}\) [km], and the maximum vehicle capacity \(\bar{x}_{i}^{h} \in \R_{>0}\) [veh]. In particular, the maximum speed allowed in the vehicle is computed as 
    \begin{equation}\label{eq:link_max_speed}
        \bar{v}_{i}^{h}(t_{k}) = v_{i, \mathrm{ff}}^{h} \left( 1 - \frac{x_{i}^{h}(t_k)}{\bar{x}_{i}^{h}}\right),
    \end{equation}
    where \(v_{i, \mathrm{ff}}^{h} \in \R_{>0}\) is the free-flow speed of the link, i.e., the speed limit. Hence, the travel time \(\tau_{i}^{h}\) is given by
    \begin{equation}\label{eq:disp_cost}
        \tau_{i}^{h}(t_{k}) = \frac{\lambda_{i}^{h}}{\bar{v}_{i}^{h}(t_{k})},
    \end{equation}
    with \(\bar{v}_{i}^{h}\) being the one in \eqref{eq:link_max_speed}. The RCSP problem \eqref{eq:rcsp} provides, for a fixed CAEV \(c\) and a given target node \(n_{\mathrm{V2G}}\), the optimal path  \(\mathcal{P}^\star_c(n_{\mathrm{V2G}})\) together with its cost \(J\big(\mathcal{P}^\star_c(n_{\mathrm{V2G}})\big)\). To this end, to determine the optimal target node for CAEV \(c\), denoted by \(n^{\star}_{c,\mathrm{V2G}}\), the RCSP is iteratively executed for each target node within the set of potential targets \(\mathcal{N}_c(t_k) =
        \left\{ \left( n_{\mathrm{V2G}}, J\big(\mathcal{P}^\star_c(n_{\mathrm{V2G}})\big)\right)\;\middle|\;
        n_{\mathrm{V2G}} \in \mathcal{V}_{\mathrm{V2G},d}
        \right\}\).
    Then, the optimal V2G target node is selected as
    \begin{equation}
        \label{eq:optimal_node}
        n^{\star}_{c,\mathrm{V2G}}(t_{k}) = \argmin_{(n,J)\in \mathcal{N}_c(t_k)} J.
    \end{equation}

    The definition of the RCSP problem \eqref{eq:rcsp} and the  choice of the target node \eqref{eq:optimal_node} represent the building block for the complete routing problem. In fact, in order to consider traffic state variations during the dispatch window \([t_{\mathrm{disp}}, t_{\mathrm{pr}}]\), the RCSP problem \eqref{eq:rcsp} is solved periodically every \(T_{\mathrm{plan}} < \Delta_{\mathrm{disp}}\) [h],  updating the travel time weights \(\tau_{i}^{h}\) in \eqref{eq:disp_cost}, for all \((i,h) \in \mathcal{E}\), according to state \(x_{i}^{h}\) of the associated link, whose dynamics is the one in \eqref{eq:link_dyn}. An high-level description of the overall procedure is reported in Algorithm \ref{alg:rcsp}.
    

    \begin{algorithm}[b]
    \caption{Periodic RCSP for CAEV path planning}
    \label{alg:rcsp}
    \begin{algorithmic}[1]
    \Require Set of CAEVs \(\bar{\mathcal{C}}_{d}\), district V2G nodes \(\mathcal{V}_{\mathrm{V2G}, d}\)
    \Ensure Optimal target node \(n^{\star}_{c, \mathrm{V2G}}\) and paths $\mathcal{P}^\star_c(n^{\star}_{c, \mathrm{V2G}})$ for all $c \in \bar{\mathcal{C}}_d$
    \State \(t_{k} = t_{\mathrm{disp}}\) 
    \Every {\(T_{\mathrm{plan}}\)} 
         \State Retrieve nodes of the CAEVs \(\bar{\mathcal{C}}_{d}\)
        \State Retrieve traffic state \(x_{i}^{h}\) \(\forall (i,h) \in \mathcal{E}\) 
        \State Compute travel time weights \(\tau_{i}^{h}(t_{k})\) as in \eqref{eq:disp_cost}
        \ForAll{\(c \in \bar{\mathcal{C}}_{d}\)}
            \State \(\mathcal{N}_{c} = \emptyset\)
            \ForAll{\(n_{\mathrm{V2G}} \in \mathcal{V}_{\mathrm{V2G}, d}\)}
            \State Compute  \(\mathcal{P}^\star_c(n_{\mathrm{V2G}})\) solving RCSP \eqref{eq:rcsp} 
            \State \(\mathcal{N}_{c} = \mathcal{N}_{c} \cup \left\{\big(n_{\mathrm{V2G}}, J(\mathcal{P}^{\star}_{c}(n_{\mathrm{V2G}}))\big)\right\}\)
            \EndFor
            \State Select the target node \(n^{\star}_{c, \mathrm{V2G}}\) as \eqref{eq:optimal_node}
            \State Define the target path for \(c\) as \(\mathcal{P}^{\star}_{c}(n^{\star}_{c,\mathrm{V2G}})\)
            
        \EndFor
        \State \(t_{k} = t_{k} + T_{\mathrm{plan}}\)
    \EndEvery

    \end{algorithmic}
    \end{algorithm}

    \begin{remark}[Choice of \(\bar{\mathcal{C}}_{d}\)] \label{rmk:initial_filter}
    Depending on the size of the urban network, the number of V2G nodes, and the cardinality of \(\bar{\mathcal{C}}_{d}\), the computational burden of Algorithm~\ref{alg:rcsp} may become significant. To reduce it, \(\bar{\mathcal{C}}_{d}\) can be selected through an offline filtering procedure instead of setting \(\bar{\mathcal{C}}_{d} = \bar{\mathcal{C}}\). Specifically, \eqref{eq:rcsp} can be solved for all \(c \in \mathcal{C}\) and \(n_{\mathrm{V2G}} \in \mathcal{V}_{\mathrm{V2G}}\) under free-flow conditions, i.e., \(x_{i}^{h}=0\) for all \((i,j)\in\mathcal{E}\), at \(t_k=t_{\mathrm{disp}}\), thus identifying a feasible subset \(\bar{\mathcal{C}}\). Then, \(\bar{\mathcal{C}}_{d}\) can be obtained by ordering \(\bar{\mathcal{C}}\) according to the cost of \eqref{eq:rcsp} and selecting the smallest subset satisfying \(\sum_{c \in \bar{\mathcal{C}}_{d}} E_{c,\mathrm{GSS}}(t_{\mathrm{disp}}) \geq \hat{E}^{\star}_{d,\mathrm{GSS}}\).
    \end{remark}

    \subsection{CAEV-level MPC scheme}
    Once the target node \(n^{\star}_{c,\mathrm{V2G}}(t_k)\) and the associated optimal path \(\mathcal{P}^{\star}_{c}(n^{\star}_{c,\mathrm{V2G}})\) are computed for each vehicle  \(c \in \bar{\mathcal{C}}_{d}\) over the horizon \([t_k, t_{\mathrm{pr}}]\), with \(t_k \geq t_{\mathrm{disp}}\), CAEV \(c\) is controlled to follow the planned path while satisfying speed, timing, and energy constraints. The vehicle position along the path is described by a 
    normalized coordinate \(p_c \in [0,1]\), defined as \( p_c = \frac{s_c}{\lambda_{c, \mathrm{tot}}}\), 
    where \(s_c \in \R_{\geq 0}\) is the traveled distance [km] along \(\mathcal{P}^{\star}_{c}(n^{\star}_{c,\mathrm{V2G}})\), while \(\lambda_{c, \mathrm{tot}} =\sum_{(i,h) \in \mathcal{P}^{\star}_{c}(n^{\star}_{c,\mathrm{V2G}})} \lambda_{i}^{h}\). Hence, \(p_c = 0\) corresponds to the departure node and \(p_c = 1\) to the target V2G node \(n^{\star}_{c,\mathrm{V2G}}\).

    Then, an MPC problem is formulated to compute the vehicle speed \(u_c \in \mathbb{R}_{\ge 0}\) for CAEV \(c\), so that it follows the planned path over the horizon \([t_k, t_{\mathrm{pr}}]\), with \(p_c(t_k)=0\), and ensuring \(p_c(t_{\mathrm{pr}})=1\) and \(E_{c,\mathrm{mob}}(t_{\mathrm{pr}}) > 0\).

    Let \(t_j \in [t_k, t_{\mathrm{pr}}]\) denote the current time instant  at which the MPC problem is solved, and let \(T_{\mathrm{MPC}} < T_{\mathrm{plan}}\) be the sampling time. Define the prediction horizon as \( N_{\mathrm{MPC}} = \frac{t_{\mathrm{pr}} - t_j}{T_{\mathrm{MPC}}}\)
    and let \(\bar j\) denote the corresponding discrete-time index. Denote by \(\mathbf{u}_c = \{ u_c(\bar j), \dots, u_c(\bar{j}+N_{\mathrm{MPC}}-1) \}\) the sequence of predicted control inputs over the horizon. Then, the MPC problem over the  window 
\([\bar j, \bar j + N_{\mathrm{MPC}}]\) is formulated as
    \begin{subequations}
       \label{eq:mpc}
        \begin{align}
            \mathbf{u}_c^\star 
            = & \argmin_{\mathbf{u}_c \in \mathbb{R}_{\ge 0}^{N_{\mathrm{MPC}}}}
            \sum_{h=\bar j}^{\bar j+N_{\mathrm{MPC}}-1} 
            \big(1 - p_c(h)\big) 
            \label{eq:mpc_cost} \\
            \text{s.t. } 
            & p_c(h+1) 
            = p_c(h) 
            + \frac{u_c(h) T_{\mathrm{MPC}}}
            {\lambda_{c,\mathrm{tot}}}
            \label{eq:mpc_position} \\
            & E_{c,\mathrm{mob}}(h+1) \text{ as in \eqref{eq:energy_dynamics}} \\
            & E_{c,\mathrm{mob}}(\bar j)
            = \hat E_{c,\mathrm{mob}}(t_j)
            \label{eq:mpc_initial_energy} \\
            & E_{c,\mathrm{mob}}(\bar j + N_{\mathrm{MPC}}) > 0
            \label{eq:mpc_terminal_energy} \\
            & p_c(\bar j) = \hat p_c(t_j)
             \quad p_c(\bar j + N_{\mathrm{MPC}}) = 1
            \label{eq:mpc_position_constraints} \\
            & 0 \le u_c(h) \le \bar v_c(t_k)
            \label{eq:mpc_input}
        \end{align}
    \end{subequations}
    where \(\bar{v}_{c} \in \R_{\geq 0}\) is the maximum speed allowed in the link in which \(c\) is located, computed during the path planning phase at instant \(t_{k}\) as in \eqref{eq:link_max_speed}, \(\hat{E}_{c,\mathrm{mob}}(t_j)\) in \eqref{eq:mpc_initial_energy} is the actual available mobility energy at time \(t_{j}\), while \(p_{c}(t_{j}) \in [0,1]\) in \eqref{eq:mpc_position_constraints} is the actual normalized position of \(c\) at instant \(t_{j}\).  Once problem \eqref{eq:mpc} is solved, the CAEV speed \(u_{c}\) is set equal to the first element of \(\mathbf{u}_c^\star\) for the time window \([t_{j}, t_{j} + T_{\mathrm{MPC}})\).

\section{Simulations and Results}\label{sec:numerical_validation}

\begin{figure}[ht]
    \centering
    \begin{subfigure}[b]{0.49\columnwidth}
        \centering
        \includegraphics[width=\columnwidth]{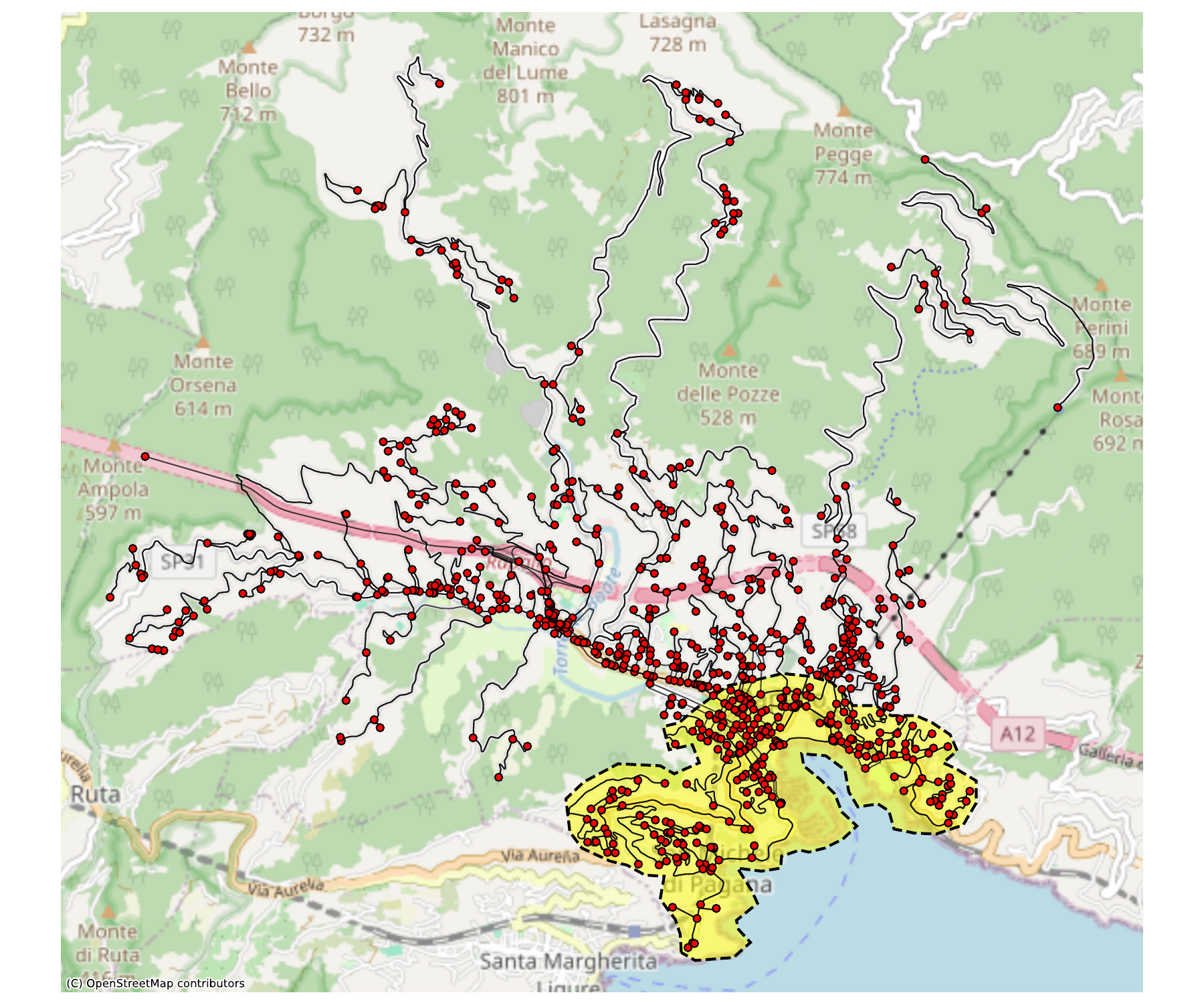}
        \caption{City map with the considered district highlighted in yellow.}
        \label{fig:rapallo_full}
    \end{subfigure}
    \hfill
    \begin{subfigure}[b]{0.49\columnwidth}
        \centering
        \includegraphics[width=\columnwidth]{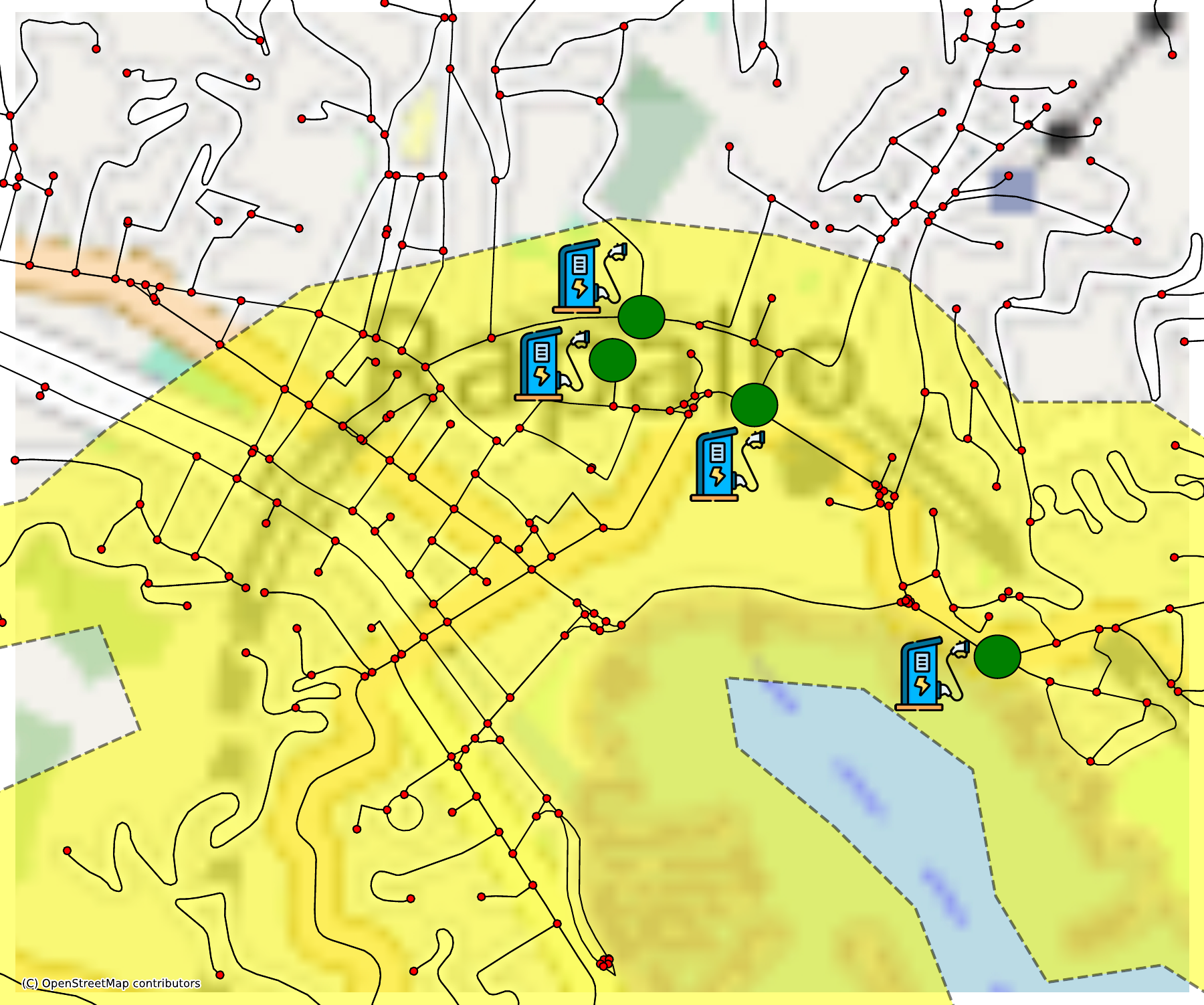}
        \caption{Zoom of the district with V2G station highlighted.}
        \label{fig:rapallo_zoom}
    \end{subfigure}
    \caption{Graph \(\mathcal{G}\) of Rapallo retrieved using the \texttt{osmnx} Python library:  district view (a) and detailed view with V2G nodes (b).}
    \label{fig:rapallo}
\end{figure}

To assess the effectiveness of the proposed architecture, simulations have been performed considering the city of Rapallo (Italy) as a case study. The urban graph, shown in Fig.~\ref{fig:rapallo_full}, has been extracted using the Python library \texttt{osmnx}. In the considered scenario, the DSO issues an energy request equal to \( \hat{E}_{1, \mathrm{GSS}}^{\star} = 200 \) kWh for a single energy district, highlighted in yellow in Fig.~\ref{fig:rapallo_full}, at time \( t_{\mathrm{disp}} = 0 \), with \( t_{\mathrm{pr}} = 10 \) min. Four V2G nodes are considered within the selected district; their locations, shown in Fig.~\ref{fig:rapallo_zoom}, reflect the position of existing charging stations.

The traffic dynamics is simulated using the model described in Section~\ref{sec:urban_traffic_dynamics}. The link lengths \( \lambda_{i}^{h} \) are obtained via \texttt{osmnx}. The inflow \( d_{i}^{h} \) and outflow \( s_{i}^{h} \) terms in \eqref{eq:link_dyn} are defined according to the link type. Let \( \mathcal{A} \subset \mathcal{V} \) denote the set of junctions corresponding to incoming extra-urban connections. Then, the inflow is defined as \( d_{i}^{h}(k) = 1500 \) if \( i \in \mathcal{A} \), and \( d_{i}^{h}(k) = 0 \) otherwise. Regarding the exit flow term in \eqref{eq:exit_flow}, it is defined through \( \epsilon_{i}^{h} = 1 \) if \( h \in \mathcal{T} \), and \( \epsilon_{i}^{h} = 0 \) otherwise. The sets \( \mathcal{A} \) and \( \mathcal{T} \) are manually defined through a graphical user interface. The saturation flow is uniformly set to \( \Phi_{i}^{h} = 1500 \) veh/h for all \( (i,h) \in \mathcal{E} \). The initial vehicle density on each link is set to 50\% of its maximum capacity \( \bar{x}_{i}^{h} \), computed from the link length assuming an average vehicle spacing of \( 7 \cdot 10^{-3} \) km. The turning rate map \( \xi_{j} \) is defined as in Remark \ref{rmk:turning_rates}, choosing the set \( \mathcal{T}_{\mathrm{sink}} \) as two terminal nodes located in the south-east area of Fig.~\ref{fig:rapallo_full}.

\begin{figure}[ht]
    \centering
    \begin{subfigure}[b]{0.57\columnwidth}
        \centering
        \includegraphics[width=\columnwidth]{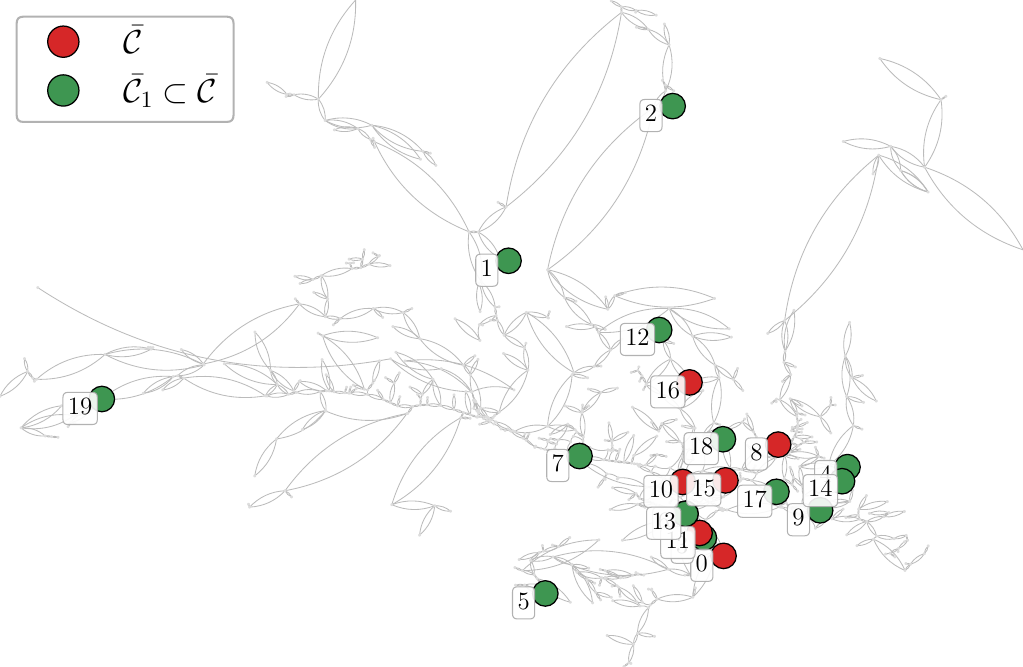}
    \end{subfigure}
    \hfill
    \begin{subfigure}[b]{0.41\columnwidth}
        \centering
        \includegraphics[width=\columnwidth]{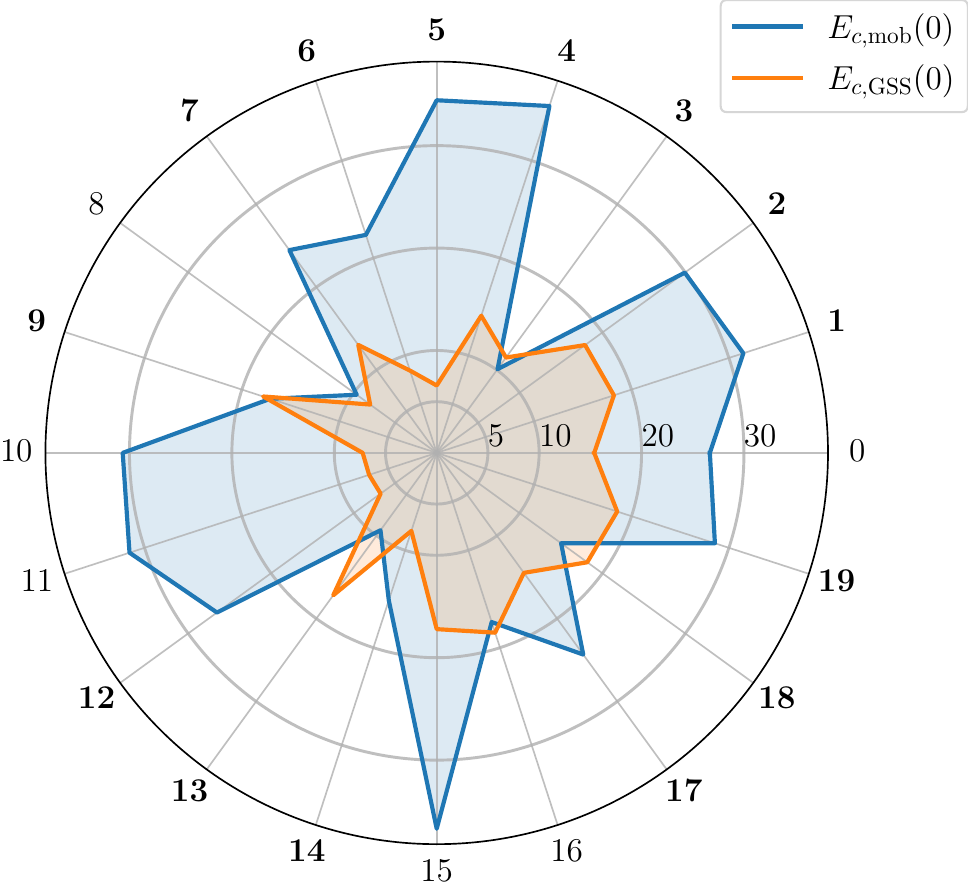}
    \end{subfigure}
    \caption{Left: simplified graph of the city showing the initial positions of all CAEVs in \(\bar{\mathcal{C}}\) (red) and of the subset \(\bar{\mathcal{C}}_{1} \subset \bar{\mathcal{C}}\) (green). The corresponding vehicle indices are also reported. Right: spider plot of the initial energy partition values for all CAEVs. Vehicle indices are reported along the circumference, while the CAEVs belonging to \(\bar{\mathcal{C}}_{1}\) are highlighted in bold.}
    \label{fig:initial_conditions}
\end{figure}

In the simulation, the set \( \bar{\mathcal{C}} \) satisfying Assumption~\ref{ass:cav_availability} consists of 20 CAEVs, each characterized by \( \bar{E}_{c,\mathrm{mob}} = 62 \) kWh and \( \bar{E}_{c,\mathrm{GSS}} = 20 \) kWh. Their initial positions are shown in the left part of Fig.~\ref{fig:initial_conditions}. The initial energy levels are randomly initialized as follows: \( E_{c,\mathrm{mob}}(0) \) is uniformly distributed between 15\% and 60\% of the mobility capacity, while \( E_{c,\mathrm{GSS}}(0) \) ranges between 25\% and 95\% of the GSS capacity. The result of the random initialization is reported in the right part of Fig. \ref{fig:initial_conditions}.  The mobility discharge dynamics in \eqref{eq:energy_dynamics} is characterized by parameters \( \eta_{1} = 0.076 \) and \( \eta_{2} = 1.35 \times 10^{-4} \).

First, the initial subset \( \bar{\mathcal{C}}_{1} \subseteq \bar{\mathcal{C}} \) is obtained according to the filtering procedure described in Remark~\ref{rmk:initial_filter}. The CAEVs in \( \bar{\mathcal{C}}_{1} \) are 14 and their initial position is highlighted by green markers in the left part of  Fig.~\ref{fig:initial_conditions}. Algorithm~\ref{alg:rcsp} is applied to this subset with a planning period \( T_{\mathrm{plan}} = 180 \) s. The RCSP problem is solved through an A$^\star$-guided label-setting algorithm with dominance pruning. Each partial path is represented by a label containing the accumulated distance, travel time, and energy consumption. 
Labels violating the time or energy constraints are discarded, while dominated labels are pruned. Route re-planning is triggered whenever a vehicle reaches its nearest intermediate node. The energy consumption rate is set to \( \varepsilon_{i}^{h} = 0.12 \lambda_{i}^{h} \). The MPC controller in \eqref{eq:mpc} operates with a sampling time \( T_{\mathrm{MPC}} = 18 \) s. The overall simulation horizon is 1 hour, with \( T = 0.00027 \) h.

\begin{figure}[t]
    \centering

    \centering
    \begin{subfigure}[b]{1\columnwidth}
        \begin{tikzpicture}[scale=0.85]
	   		\node [ align=center, inner sep = 0pt ]  () at (0,0){ \includegraphics[width=\columnwidth]{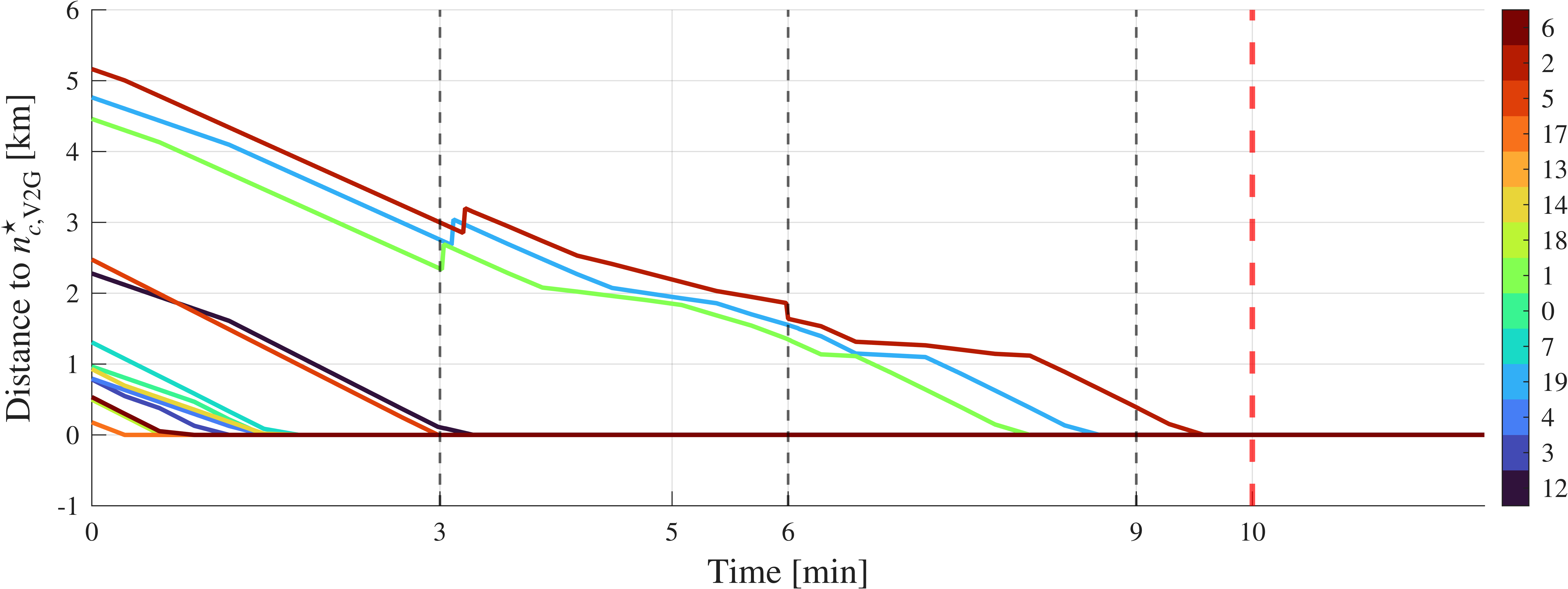}};
            \draw[thick] (-2.25,0.7) rectangle (-1.5,0); 
             \node[draw=none, font=\tiny]  at (-1.8,0.85) {Path change};
            
        \end{tikzpicture}
        
        \label{fig:cavs_distance_target}
    \end{subfigure}
    \hfill
    \begin{subfigure}[b]{1\columnwidth}
        \centering
        \includegraphics[width=\columnwidth]{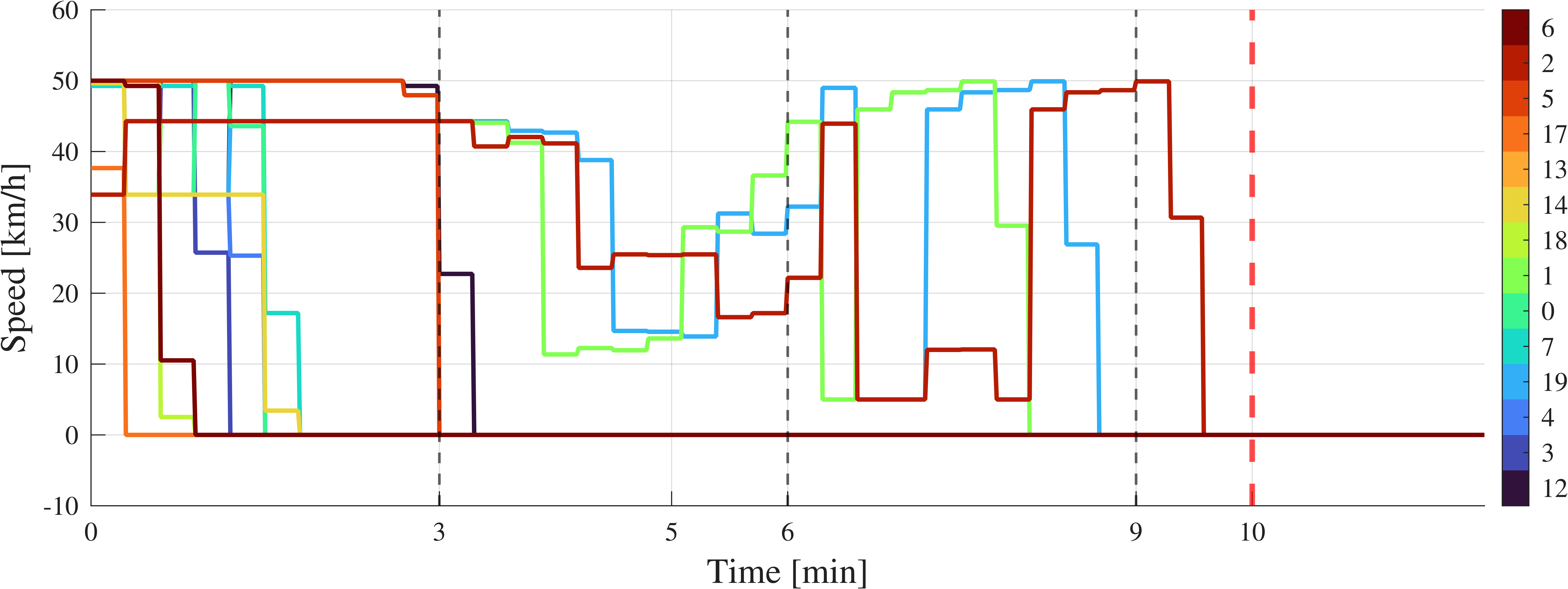}
        \label{fig:cavs_speed}
    \end{subfigure}
    \caption{Time behavior of the CAEVs distance from their respective V2G target node (top) and their speed (bottom) during the simulation. Vertical black dashed lines denote the instant in which re-planning is performed, while the vertical red dashed line is \(t_{\mathrm{pr}}\).} 
    \label{fig:cavs_results}
\end{figure}

\begin{figure}[t]
    \centering
    \begin{subfigure}[b]{0.49\columnwidth}
        \centering
        \includegraphics[width=\columnwidth]{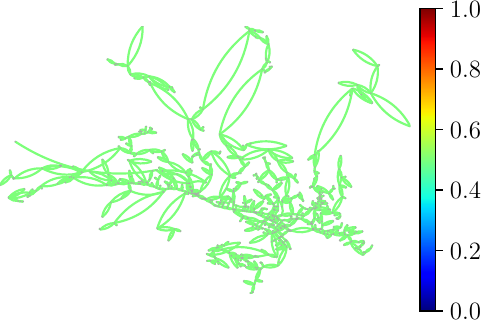}
        \label{fig:snap0_traffic}
    \end{subfigure}
    \hfill
    \begin{subfigure}[b]{0.49\columnwidth}
        \centering
        \includegraphics[width=\columnwidth]{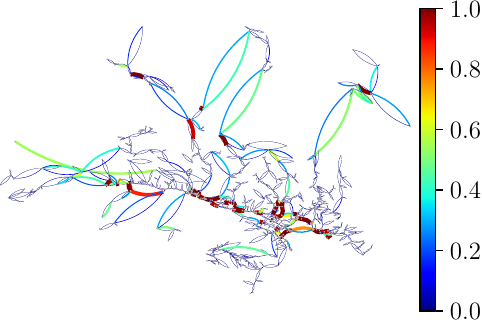}
        \label{fig:snap180_traffic}
    \end{subfigure}
    \\
    \begin{subfigure}[b]{0.49\columnwidth}
        \centering
        \includegraphics[width=\columnwidth]{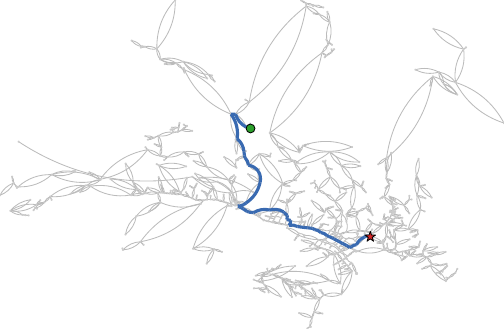}
        \label{fig:snap0_path}
    \end{subfigure}
    \hfill
    \begin{subfigure}[b]{0.49\columnwidth}
        \centering
        \includegraphics[width=\columnwidth]{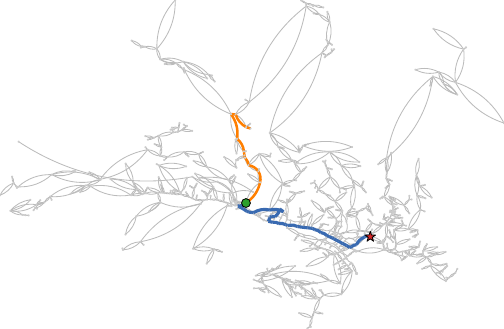}
        \label{fig:snap180_path}
    \end{subfigure}
    \caption{
    Traffic state and planned path at the first two re-planning instants. Top row: occupancy ratio $x_{i}^{h}/\bar{x}_{i}^{h}$ on each edge $(i,h)\in\mathcal{E}$ at $t_{0}=0$\,s (left) and $t_{1}=180$\,s (right), encoded by both colour and line width. Bottom row: corresponding path assigned to the CAEV 1 by Algorithm~\ref{alg:rcsp}, where the blue edges indicate the planned route, the orange edges the already traveled one, the green marker denotes the vehicle position at the planning instant, and the star denotes the selected V2G node $n_{\mathrm{V2G},1}^{\star}$.}
    \label{fig:planning_results}
\end{figure}

The simulation results are reported in Fig.~\ref{fig:cavs_results}, which shows the time evolution of each CAEV speed and distance from its assigned V2G target node. Each dispatched vehicle \(c \in \mathcal{C}_{1}\) reaches its target before the deadline \( t_{\mathrm{pr}}\), hence it delivers the energy \(E_{c, \mathrm{GSS}}(0)\) reserved for GSS, reported in orange in the right part of Fig.~\ref{fig:initial_conditions}. It can be observed that three CAEVs, namely 1, 2, and 19, do not reach their targets before the first re-planning instant. For these vehicles, the planned path is modified, as indicated by the positive distance spike occurring immediately after re-planning. This behavior is due to  variations in the traffic state, as illustrated in Fig.~\ref{fig:planning_results}.

    \section{Conclusion}\label{sec:conclusion}
   This paper presented an energy-mobility dispatch framework for GSS provision in smart cities. Unlike conventional V2G approaches, the proposed method jointly accounts for traffic dynamics, battery energy management, and grid requirements. VBP preserves driving functionality while enabling reliable grid support. A prescribed energy request from the DSO is satisfied by periodically solving a congestion-aware RCSP over the urban network and regulating vehicle speed through an MPC strategy subject to time and mobility energy constraints. Simulations based on the urban network of Rapallo (Italy), modeled through a macroscopic traffic framework, validate the effectiveness of the proposed approach under dynamic congestion conditions. Future work will address the integration of traffic prediction models, the inclusion of CAEV mobility tasks within both the planning and control layers, and the management of multiple requests from different energy districts.
	
	\bibliographystyle{IEEEtran}
	\bibliography{IEEEabrv,biblio}

@article{rua2026smart,
  title={Smart cities and open innovation: what is known, how it is known and future agenda},
  author={Rua, Orlando Lima and Arias-Oliva, Mario and Shu, Ziwei and Souto-Romero, Mar},
  journal={Discover Sustainability},
  year={2026},
  publisher={Springer}
}

@article{zhang2020joint,
  title={Joint optimal power flow routing and vehicle-to-grid scheduling: Theory and algorithms},
  author={Zhang, Shiyao and Leung, Ka-Cheong},
  journal={IEEE Transactions on Intelligent Transportation Systems},
  volume={23},
  number={1},
  pages={499--512},
  year={2020},
  publisher={IEEE}
}

@article{hu2021distributed,
  title={A distributed MPC to exploit reactive power V2G for real-time voltage regulation in distribution networks},
  author={Hu, Jindi and Ye, Chengjin and Ding, Yi and Tang, Jinjiang and Liu, Si},
  journal={IEEE Transactions on Smart Grid},
  volume={13},
  number={1},
  pages={576--588},
  year={2021},
  publisher={IEEE}
}

@article{dong2023multi,
  title={Multi-agent reinforcement learning for intelligent V2G integration in future transportation systems},
  author={Dong, Jiawei and Yassine, Abdulsalam and Armitage, Andy and Hossain, M Shamim},
  journal={IEEE transactions on intelligent transportation systems},
  volume={24},
  number={12},
  pages={15974--15983},
  year={2023},
  publisher={IEEE}
}

@article{wang2022coordinated,
  title={Coordinated electric vehicle active and reactive power control for active distribution networks},
  author={Wang, Yi and Qiu, Dawei and Strbac, Goran and Gao, Zhiwei},
  journal={IEEE Transactions on Industrial Informatics},
  volume={19},
  number={2},
  pages={1611--1622},
  year={2022},
  publisher={IEEE}
}

@article{abdulaal2016solving,
  title={Solving the multivariant EV routing problem incorporating V2G and G2V options},
  author={Abdulaal, Ahmed and Cintuglu, Mehmet H and Asfour, Shihab and Mohammed, Osama A},
  journal={IEEE Transactions on Transportation Electrification},
  volume={3},
  number={1},
  pages={238--248},
  year={2016},
  publisher={IEEE}
}

@article{mavlutova2023urban,
  title={Urban transportation concept and sustainable urban mobility in smart cities: a review},
  author={Mavlutova, Inese and Atstaja, Dzintra and Grasis, Janis and Kuzmina, Jekaterina and Uvarova, Inga and Roga, Dagnija},
  journal={Energies},
  volume={16},
  number={8},
  pages={3585},
  year={2023},
  publisher={MDPI}
}

@article{vstogl2024electric,
  title={Electric vehicles as facilitators of grid stability and flexibility: A multidisciplinary overview},
  author={{\v{S}}togl, Ond{\v{r}}ej and Miltner, Marek and Zanocco, Chad and Traverso, Marzia and Star{\`y}, Old{\v{r}}ich},
  journal={Wiley Interdisciplinary Reviews: Energy and Environment},
  volume={13},
  number={5},
  pages={e536},
  year={2024},
  publisher={Wiley Online Library}
}

@article{torkey2024transportation,
  title={Transportation electrification: a critical review of EVs mobility during disruptive events},
  author={Torkey, Alaa and Zaki, Mohamed H and El Damatty, Ashraf A},
  journal={Transportation Research Part D: Transport and Environment},
  volume={128},
  pages={104103},
  year={2024},
  publisher={Elsevier}
}

@article{bonfiglio2025vehicle,
  title={Vehicle-to-home service via electric vehicle energy storage virtual partitioning},
  author={Bonfiglio, Andrea and Minetti, Manuela and Procopio, Renato},
  journal={IEEE Transactions on Industry Applications},
  year={2025},
  publisher={IEEE}
}

@article{de2025exploring,
  title={Exploring complementary effects of solar and wind power generation},
  author={de Andrade Melo, Gustavo and Oliveira, Fernando Luiz Cyrino and Ma{\c{c}}aira, Paula Medina and Meira, Erick},
  journal={Renewable and Sustainable Energy Reviews},
  volume={209},
  pages={115139},
  year={2025},
  publisher={Elsevier}
}

@book{twidell2021renewable,
  title={Renewable energy resources},
  author={Twidell, John},
  year={2021},
  publisher={Routledge}
}

@article{akorede2010distributed,
  title={Distributed energy resources and benefits to the environment},
  author={Akorede, Mudathir Funsho and Hizam, Hashim and Pouresmaeil, Edris},
  journal={Renewable and sustainable energy reviews},
  volume={14},
  number={2},
  pages={724--734},
  year={2010},
  publisher={Elsevier}
}

@article{dinopoulou2006applications,
  title={Applications of the urban traffic control strategy TUC},
  author={Dinopoulou, Vaya and Diakaki, Christina and Papageorgiou, Markos},
  journal={European Journal of Operational Research},
  volume={175},
  number={3},
  pages={1652--1665},
  year={2006},
  publisher={Elsevier}
}

@book{rajamani2006vehicle,
  title={Vehicle dynamics and control},
  author={Rajamani, Rajesh},
  year={2006},
  publisher={Springer}
}

@article{zanella2014internet,
  title={Internet of things for smart cities},
  author={Zanella, Andrea and Bui, Nicola and Castellani, Angelo and Vangelista, Lorenzo and Zorzi, Michele},
  journal={IEEE Internet of Things journal},
  volume={1},
  number={1},
  pages={22--32},
  year={2014},
  publisher={Ieee}
}

@article{bibri2017smart,
  title={Smart sustainable cities of the future: An extensive interdisciplinary literature review},
  author={Bibri, Simon Elias and Krogstie, John},
  journal={Sustainable cities and society},
  volume={31},
  pages={183--212},
  year={2017},
  publisher={Elsevier}
}

@article{jia2022control,
  title={Control for smart systems: Challenges and trends in smart cities},
  author={Jia, Qing-Shan and Panetto, Herv{\'e} and Macchi, Marco and Siri, Silvia and Weichhart, Georg and Xu, Zhanbo},
  journal={Annual Reviews in Control},
  volume={53},
  pages={358--369},
  year={2022},
  publisher={Elsevier}
}
    \balance
\end{document}